\documentclass[pre, twocolumn, amsmath, amssymb, superscriptaddress]{revtex4}

\usepackage{graphicx}
\usepackage{dcolumn}
\usepackage{bm}
\usepackage{braket}
\usepackage{xcolor}
\usepackage{booktabs}
\usepackage{multirow}
\usepackage{comment}

\usepackage{amsmath}

\newcommand{\beq}{\begin{equation}}
\newcommand{\eeq}{\end{equation}}

\begin{document}

\title{Probing leukemia cells behavior under starvation}

\author{Simone Scalise}
\affiliation{Center for Life Nano \& Neuro Science, Istituto Italiano di Tecnologia, Viale Regina Elena 291,  00161, Rome, Italy}

\affiliation{Department of Physics, Sapienza University, Piazzale Aldo Moro 5, 00185, Rome, Italy}



\author{Giorgio Gosti}
\affiliation{Istituto di Scienze del Patrimonio Culturale, Consiglio Nazionale delle Ricerche,00010, Montelibretti, Italy}

\author{Giancarlo Ruocco}
\affiliation{Center for Life Nano \& Neuro Science, Istituto Italiano di Tecnologia, Viale Regina Elena 291,  00161, Rome, Italy}

\affiliation{Department of Physics, Sapienza University, Piazzale Aldo Moro 5, 00185, Rome, Italy}

\author{Giovanna Peruzzi }
\affiliation{Center for Life Nano \& Neuro Science, Istituto Italiano di Tecnologia, Viale Regina Elena 291,  00161, Rome, Italy}

\author{Mattia Miotto  \footnote{\label{corr} For correspondence write to: mattia.miotto@roma1.infn.it}}
\affiliation{Center for Life Nano \& Neuro Science, Istituto Italiano di Tecnologia, Viale Regina Elena 291,  00161, Rome, Italy}

\begin{abstract}
The ability of a cancer cell population to achieve heterogeneity in their phenotype distributions offers advantages in tumor invasiveness and drug resistance. Studying the mechanisms behind such observed heterogeneity in mammalian cells presents challenges due for instance to the prolonged proliferation times compared to widely studied unicellular organisms like bacteria and yeast.
Here, we studied the response of leukemia cell populations to serum starvation via a protocol, we recently developed,  that makes use of live cell fluorescence and flow cytometry in combination with a quantitative analytical model to follow the population proliferation while monitoring the dynamics of its phenotype distributions. 
We found that upon switching between a serum-rich to a serum-poor media, leukemia cells (i) maintain a memory of the previous environment up to one generation even in the presence of severe medium-depletion, before (ii)  adapting their growth and division rates to the novel environment while preserving a sizer-like division strategy. 
Finally, looking at the mitochondria content of the proliferating vs non-proliferating cells, we found that the latter is characterized by a higher number of older mitochondria, suggesting a possible functional role of the observed asymmetric partitioning of (aged) mitochondria in leukemia cells.  
\end{abstract}

\maketitle

\section{Introduction}
Adaptation to environmental changes is one of the driving forces of evolution~\cite{hoffmann1997extreme}. At the level of unicellular organisms, like bacteria, the environment such organisms live in can fluctuate in timescales comparable with the division time of the single cell~\cite{expexp}. This vouches for the cells to evolve mechanisms to withstand/recover from sudden changes in the environments (e.g. temperature, pH, or nutrient levels changes) both at the level of the single cell and at that of the population. 
Examples are the presence of heat shock proteins, a class of proteins specialized in protecting cells from sudden thermal stresses, the evolution of proteins in extremophiles, which display overall higher thermal resistance~\cite{Miotto2020diff}, or the presence of different molecular circuits to process distinct nutrient sources~\cite{Wilson2010}. Bacteria for instance can extract energy from the digestion of several different sugars, although with different efficiencies~\cite{Wilson2010, Sekar2020}.
Indeed, among the different sources of environmental change/stress, starvation assumes a peculiar place due to the direct link between nutrient availability and the growth law displayed by the cell population~\cite{EnricoBena2021, Kellogg2022}.

Previous works showed a nontrivial link between the availability of nutrients and the growth of the population. In particular, at a phenomenological level different growth laws have been introduced on the basis of the relationship between cell size at growth and division.  Three strategies (and combinations thereof) are often defined depending on whether cells divide after reaching a certain target size (sizer model), after having added a determined volume (adder model), or after having grown for a specific time (timer regime). 
The determination of the preferred growth model for bacterial cell populations has been made possible by the recent advances in live cell imaging techniques.  

In general, an extension of the obtained results to mammalian cells is still hindered by the difficulties of monitoring the proliferation for timescales long enough to observe multiple division events.  Some progress in this respect has been made by the work of  Kafri ~\textit{et al.}~\cite{Kafri2013}, who used fluorescence microscopy to get dynamical information from fixed cells and reveal feedback linking between cell growth and cell cycle; or one of Sung and coworkers, who developed a phase microscopy technique for highly accurate measurements of cell mass in adherent mammalian cells~\cite{Sung2013}.

Here, we performed a series of proliferation experiments under different degrees of serum starvation on a leukemic cell line (Jurkat T-cell). In particular, we applied a protocol we recently developed to monitor the proliferation of cell population via flow kinetic experiments of properly marked cells combined with a minimal theoretical model that allows for the quantification of the growth and division statistics of the population~\cite{peruzzi2021asymmetric, miotto2023determining}. Following the growth of a population in standard growing conditions, we showed how progressive dilution of a cytoplasmic dye permits the stratification of the whole population by the generation each cell belongs to at a given time; and that the intensity of the forward scattering (FSC) provides a proxy for the cell size. Applying our protocol to leukemia T cells (Jurkat), we found that Jurkat cell proliferation curves are reproduced assuming that cell growth and division rates are power functions of the cell size~\cite{miotto2023determining}. 
Performing experiments with different serum amounts (starvation conditions), we found that (i) the division strategy is preserved, although (ii) cells adapt growth and division rates in a serum-dependent manner. Moreover, upon switching the growing conditions from standard to starved ones, cells maintain a memory of the previous conditions, continuing proliferating at the same pace as in the previous environment for a time inversely proportional to the difference between the environments.

\section{Results}
We analyze the effects that serum starvation has on leukemia cell proliferation and whether and how it reflects at the level of cell phenotypic distributions and their dynamics. 
To this aim, we deployed an experimental protocol based on flow cytometry measurements, originally developed to study the partition noise of internal components during cell division \cite{peruzzi2021asymmetric} and recently extended to determine the division strategy of leukemic cells \cite{miotto2023determining}.

\subsection{Monitoring population dynamics under  starvation via flow cytometry measurements}
\begin{figure*}
    \centering
    \includegraphics[width=\textwidth]{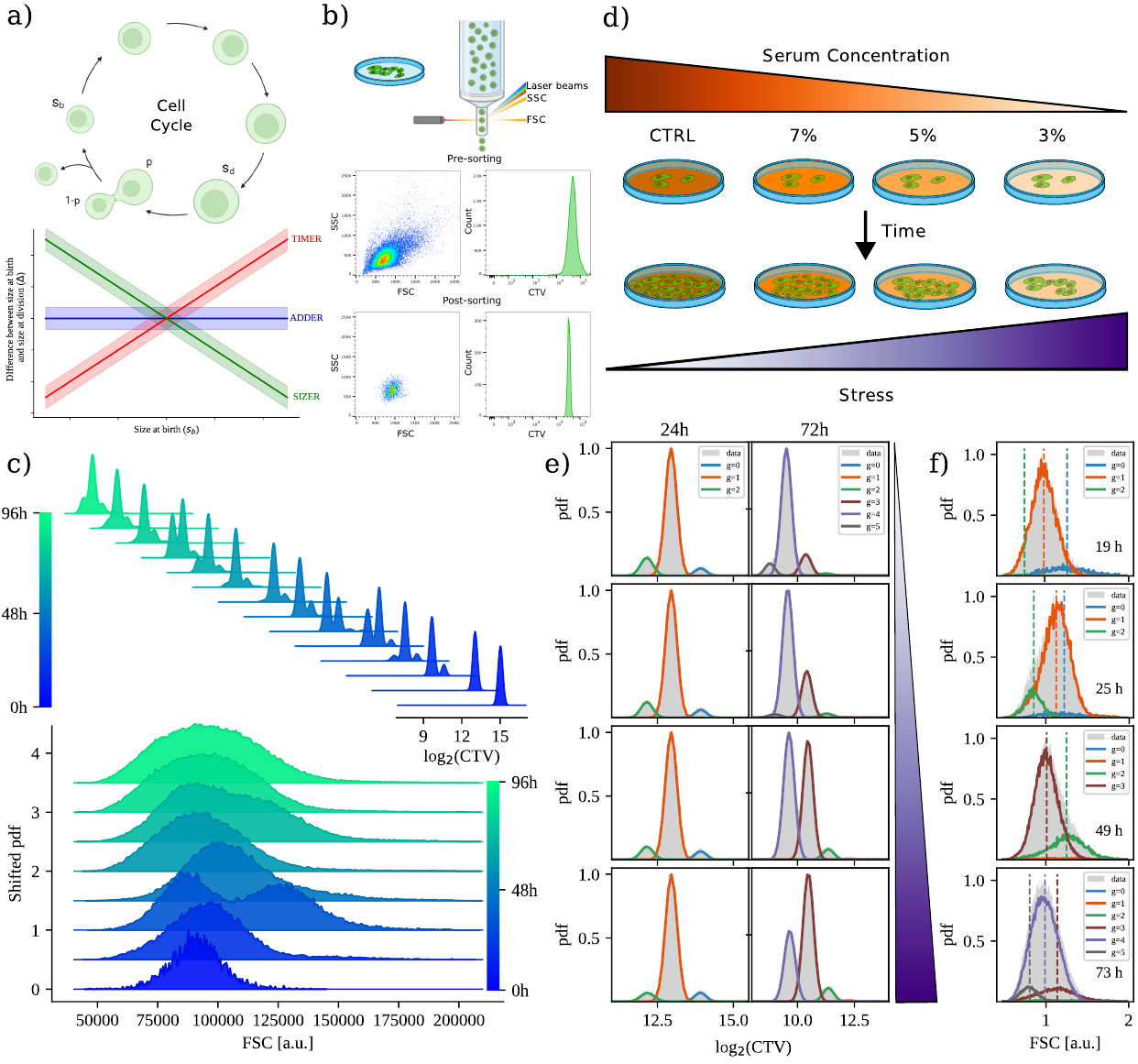}
    \caption{\textbf{Experimental protocol to follow cell proliferation under starvation.} \textbf{a)} (top) Schematic representation of cell growth and division. A mother cell with starting size $s_b$ grows up to a size $s_d$ and then splits into two daughter cells whose starting sizes are fractions of the mother cell size. 
    (bottom) Behaviors of the three size-homeostasis models.  The different models can be distinguished based on the correlation they present between the size at birth $s_b$ and how much they have grown, given by the difference between the size at division and at birth $\Delta = s_d-s_b$.
    \textbf{b)} Schematic representation of the time course protocol: at different time points an aliquot of cells is collected from each different well and analyzed via a flow cytometer that collects both FSC, SSC, and CTV intensities for each analyzed cell.\textbf{d1)} SSC vs FSC intensities together with the respective distribution of CTV (CellTrace Violet, cytoplasm marker) intensity for the initial (Pre-sorting) population of marked cells.
    The bottom panels display the SSC vs FSC intensities and distribution of CTV  intensity for cells with the established collection gate (Post-sorting).
    \textbf{c)} (top) Time evolution of the population CTV fluorescence intensity. Colors from blue to green represent different time points along the experimental time course, from time zero (bottom-right) to 96 hours (top-left). 
    (bottom)  Density distribution of the forward scattering intensity of the cell population at different times. Colors range from blue to green as the time goes from zero (sorting of cells, i.e. start of the experiment) to 96 hours.   
    \textbf{d)} Serum-starvation protocol: the initial sorted population is divided and kept in culture in four different wells each with different serum concentrations media (Ctrl, 7\%, 5\%, and 3\%) for the whole experiment (see Methods). Proliferation is influenced by serum concentration.
    \textbf{e)} Density distribution (grey curves) and best fit of a Gaussian Mixture Model (colored curves) of the CTV fluorescence intensity measured at different times during proliferation for the different starvation conditions. The first and second columns are respectively snapshots at 24 and 72 hours from the sorting and culturing process while from top to bottom, the serum concentration decreases (Ctrl, 7\%, 5\% and 3\%). Curves colored from blue to grey are ordered according to the identified generations.
    \textbf{f)} Density distribution and best fit of the measured FSC intensities for a specific growth condition (Ctrl) at different times. Colored curves highlight the subpopulations corresponding to the different generations that have been identified thanks to the Gaussian Mixture model fitting of the CTV intensities. Intensities have been rescaled by a factor of $10^5$. The vertical dotted lines indicate the mean value of the distribution of each generation.} 
    \label{fig:1}
\end{figure*}

To follow cell proliferation via flow cytometry measurements, we followed the protocol described in Miotto~\textit{et al.}~\cite{miotto2023determining}. In brief, we stained cells with the cytoplasmic fluorescent dye, CellTrace\textsuperscript{\texttrademark} Violet (CTV) and isolated, via cell sorter,  an initial subpopulation characterized by a narrow CTV distribution centered around the maximum the of the dye distribution of the whole population (see Figure~\ref{fig:1}b).
Culturing sorted cells in standard growing conditions (see Methods), it is possible to track cell dynamics over multiple cell generations by analyzing the fluorescence intensity dilution over time. In particular, we took samples at different times and acquired the CTV and Forward Scattering (FSC) intensities for each sampled cell.  Figure~\ref{fig:1}c displays the time evolution of the CTV (top) and FSC (bottom) intensities.
Comparing the CTV distributions at different time acquisition, a progressive shift of the initial fluorescence distribution (blue curve in the low-right corner of the panel) can be seen with the appearance of new peaks in subsequent data points.
Such peaks represent newly forming generations. Indeed, CTV binds uniformly to cytoplasmic proteins. 
As, after each cell division, CTV is partitioned between the two daughter cells, the average fluorescence of the first generation of divided cells (considering the sorted cells as mother cells or generation zero cells) is half of the mean of the previous generation.
 Starting with a sufficiently narrow initial distribution, the generation peaks can be distinguished. 
Note that this method allows for the tracking of several generations.

To characterize what happens when growth conditions are altered, we cultured the sorted cells in different growth conditions: as explained in more detail in Methods, we split the sorted population into four separate wells, each with the same media but with different levels of nutrient concentration (see Figure~\ref{fig:1}d for a sketch).  Standard growth conditions are obtained with medium supplemented with 10\% serum, while starved ones are reproduced decreasing the amount of serum, i.e. medium supplemented with 7\%, 5\%, and 3\% serum respectively. 

In Figure \ref{fig:1}e, we reported the distribution of CTV intensities for two different time points, taken respectively 24h and 72h after the initial population is sorted. In grey, the distributions of the CTV intensity in the log2 base are shown.  The four rows correspond to the different serum concentrations at which the cells are cultured, with the level of starvation increasing from top (control growth) to bottom (3\% serum growth). Focusing on the two columns, it is possible to notice the proliferation difference among the four growth conditions types. 

To quantify the observed differences, we applied a fitting protocol via a Gaussian Mixture Model of the form: $P(\ln x) = \sum_g w_g N(\ln x, \bar x_g, \sigma_g)$ combined with an Expectation Maximization algorithm (see Methods for details), to identify the different generations from the CTV intensity profiles. Colored lines represent the different identified generations, i.e. the Gaussian distribution obtained as the best fit for the experimental data.
Comparing the same generations at different times/levels of serum allows us to appreciate the effect of starvation:
 if at 24h all four conditions show roughly the same level of proliferation (in all of them the highest peak corresponds to the same generation), at 72 h, the differences are more evident: cells grown under normal conditions show a peak at the fourth generation (violet curve) with the fifth already forming; decreasing in concentration, proliferation is further and further behind, until cells grown in 3\% serum medium where the highest peak is given by the third generation (brown curve) and the fourth is forming.
This difference in proliferation confirms what we expected: the decrease in serum within the growth medium leads to a progressive slowing down in proliferation. 

Along with the changes in CTV intensity, we monitor the changes in FSC distribution. As shown in Figure~\ref{fig:1}c, the intensities increase overall, accompanied by an increase in variance, fluctuating from the initial (blue curve) to higher values in the last acquisition points. 
This is because the initial population is sorted according to CTV intensity values, which leads to an out-of-equilibrium size distribution.

Thanks to the fact that for each cell we measure the CTV and FSC signals at each time point, we can use the information obtained with the fitting by GM to identify for the FSC distributions the subpopulations corresponding to the different generations. 
This allows us to decompose a given FSC distribution into as many sub-distributions as there are generations identified at that specific time point, as reported in Figure~\ref{fig:1}f for four different times of the same population (we reported the 10\% serum condition as an example).  
By comparing the distributions of the same generation in different snapshots, it can be seen that the newer generations have distributions shifted towards smaller FSC values than the older ones, as also evidenced by the mean values of the individual distributions (dashed lines).

\subsection{Growth rate changes after an adaptation time}

To get insights into the effect of serum deprivation on the cells, we started analyzing the behavior at the population level.
Along the lines of Miotto \textit{et al.}~\cite{miotto2023determining}, we sought a model for the growth and division dynamics of a population of cells able to describe the behavior of the cell size distribution in time (experimentally captured by the FSC intensity). In addition, we can use CTV data to stratify the cell population in terms of the cell generations; thus we looked for an analytical formulation from the quantity, $\rho_g(s, t)$, the probability density function of finding cells with size, s, at time, t, and generation, g. 

To do so, we start considering $n_g(s(t),t)$ the number of cells belonging to generation $g$ and having size $s$ at time $t$. In balanced growth conditions, it will evolve in time following the equation:

\begin{multline}
\label{eq:PBE}
\frac{\partial n_g(s,t)}{\partial t} + \frac{\partial \left(g(s) \cdot n_g(s,t)\right)}{\partial s} = -\gamma(s) n_g(s(t),t) +  \\ + 2 \int_0^{\infty} d\eta~ \gamma(\eta)~\phi(s|p\eta)~n_{g-1}(\eta,t)
\end{multline}

where $g(s)$ and $\gamma(s)$ are the size-dependent growth and division rate, respectively, and $\phi(x|py)$  quantifies the probability that a daughter cell inherits a fraction p of mother cell size y (see \cite{miotto2023determining} for a complete discussion).

In particular, we assumed that both growth and division rates can be written as powers of cell size~\cite{Osella2014, NietoAcuna2019, Totis2021, Jia2021}
\begin{align}
\label{eq:sizepow}
    g(s) &= \frac{ds}{dt}= \lambda s^\alpha & \gamma(s) &= \kappa s^\beta
\end{align}
as this choice allows to recover of all main size-homeostatic strategies, i.e. timer, sizer, or added regime by tuning the parameter $\omega = \beta -\alpha$ (see \cite{Nieto2020}). 

We now introduce the probability of finding a cell with size $s$ at time $t$ and generation $g$
\beq
\label{eq:prob}
\rho_g(s,t) = \frac{n_g(s,t)}{N_g(t)}.
\eeq
and the fraction of cells that belong to a certain generation at each time:

\beq
P_g(t) = \frac{N_g(t)}{\sum_q N_q(t) }
\eeq

After some calculations (see \cite{miotto2023determining} for the explicit derivation), one can get an expression   for the dynamics of the fractions:

\begin{multline}
\label{eq:fractions}
\dot{P_g} = -k\braket{s^\beta}_g P_g + 2k\braket{s^\beta}_{g-1} P_{g-1} + \\+ P_g \sum_q  \left( k\braket{s^\beta}_q P_q - 2k\braket{s^\beta}_{q-1} P_{q-1} \right)
\end{multline}

 where $ <f>_g = \int ds f(s) P(s)_g$ and the each moment evolves as:

\begin{multline}
\label{eq:moments}
\braket{\dot{s}^i}_g =  \lambda \cdot i\cdot \braket{s^{(\alpha + i -1)}}_g - \Phi_g\braket{s^i}_g  -k\braket{s^{(\beta + i)}}_g + \\+ 2~k\left< p^i \right>_\pi \braket{s^{(\beta+ i)}}_{g-1}  \frac{P_{g-1}}{P_{g}}
\end{multline}

where $<p^i>_\pi$ refers to the i-th moment of $\pi(p)$. 

This model produces division times with exponential distributions \cite{NietoAcuna2019}. To recover the Erlang distributed inter-division time reported in literature \cite{Yates2017, NietoAcuna2019},  it is sufficient to introduce a series of intermediate states into which a cell has to pass before it can divide. From a biological point of view, such states can be linked to the phases of the cell cycle (see Figure~\ref{fig:1}a). Since the duration of each state is characterized by an exponential distribution of times, it is sufficient to repeat all the calculations made so far by simply introducing an additional index that considers the intermediate state the cell is in.

By doing so, we obtain an evolution equation for the moment distributions:

\begin{multline}
\label{eq:fractions_q}
\braket{\dot{s}^i}_{g,q} =  \lambda \cdot i\cdot \braket{s^{(\alpha + i -1)}}_{g,q} - \Phi_{g,q}\braket{s^i}_{g,q}  -k\braket{s^{(\beta + i)}}_{g,q} +  \\ + 
k\left< 2~p^i\right>_\pi^{\delta_{q,0}} \braket{s^{(\beta+ i)}}_{g,q -1}  \frac{P_{g, q -1}}{P_{g,q}}
\end{multline}

where now $\braket{\cdot}_{g,q}$ stands for the statistical average over $\rho_{g,q}$,   the density of cells that divided $g$ times and passed $q$-th out of Q intermediate states and $\delta_{i,j}$ is the Kronecker delta.
Likewise, the cell fractions at generation $g$ and state $q$ evolve in time following 

\begin{multline}
\label{eq:moments_0}
\dot{P_{g,q}} =  -k\braket{s^\beta}_{g,q} P_{g,q} + 2k\braket{s^\beta}_{g,q-1} P_{g,q-1} + \\ + P_{g,q} \sum_{h,w}  \left( k\braket{s^\beta}_{h,w} P_{h,w} - 2^{\delta_{w,0}}k\braket{s^\beta}_{h,  w -1} P_{h, w -1} \right)
\end{multline}

for  $q > 0$, and  

\begin{multline}
\label{eq:moments_q}
\dot{P_{g,0}} =  -k\braket{s^\beta}_{g,0} P_{g,0} + k\braket{s^\beta}_{g-1,0} P_{g-1,0} + \\ +  P_{g,0} \sum_{h,w}  \left( k\braket{s^\beta}_{h,w} P_{h,w} - 2^{\delta_{w,0}}k\braket{s^\beta}_{h, w-1} P_{h, w-1} \right)  \end{multline}

otherwise.
Although the equations just found fully describe the dynamics of the cell population, they do not constitute a closed set of equations as they may contain higher-order moments that depend on the parameter values used.
To solve the system, a moment closure strategy was adopted whereby the size distributions of the individual generations have normal moments. This choice, based on observations made on the FSC distributions divided by generations \cite{Totis2021}, was validated in Miotto \textit{et al.}~\cite{miotto2023determining} where the solutions of the differential equations were compared with the results of an agent-based stochastic simulation.  
Such simulation, where an initial population of cells grow and divide following the same grow and division rates functional form used in Eq.~\ref{eq:PBE}, reproduces the different size control mechanisms upon varying the $\omega = \beta -\alpha$ parameter.

\begin{figure*}
    \centering
    \includegraphics[width=\textwidth]{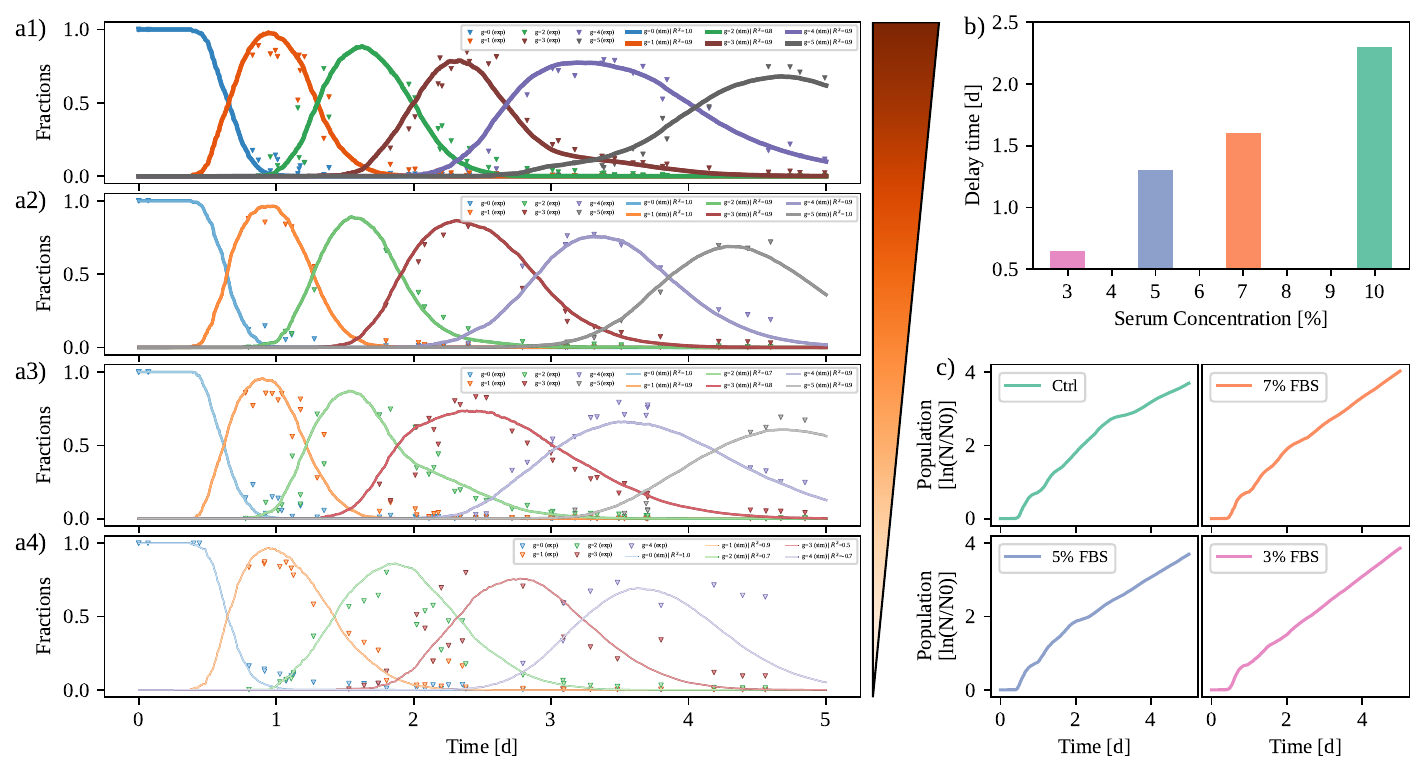}
    \caption{\textbf{Population dynamics vs serum starvation levels.} 
    \textbf{a1-a4)} Measured fractions of cells in different generations as a function of time (dots) and curves given by the best fit of the minimal model, described by Eqs.~\ref{eq:moments} and Eqs.~\ref{eq:fractions}. From top to bottom, the serum concentration decreases (Ctrl, 7\%, 5\%, and 3\%).
    \textbf{b)} Time delay at which cells change growth and division rates depending on the different levels of serum concentration in the medium. 
    \textbf{c)} Growth curves of cell populations for the different levels of nutrient deprivation.
    }
    \label{fig:2}
\end{figure*}

\begin{table}[htbp]
\centering
\caption{Values of parameters}
\label{tab:parameters}
\begin{tabular}{cccccc}

                       &  & CTRL & 7\% & 5\% & 3\% \\ \cline{2-6}
\multirow{6}{*}{\rotatebox[origin=c]{90}{I set}} &  $\alpha$  & 1 & 1 & 1 & 1 \\
                       &  $\beta$   & 5 & 5 & 5 & 5 \\
                       &  Q         & 5 & 5 & 5 & 5 \\
                       &  $\lambda$ & 1.0 & 1.05 & 1.1 & 1.0 \\
                       &  $\kappa$  & 0.0769 & 0.0769 & 0.0846 & 0.0769 \\
                       &  f         & 13 & 13 & 13 & 13 \\ \hline
\multirow{6}{*}{\rotatebox[origin=c]{90}{II set}}      &  $\alpha$  & 1 & 1 & 1 & 1 \\
                       &  $\beta$   & 5 & 5 & 5 & 5 \\
                       &  Q         & 6 & 6 & 6 & 6 \\
                       &  $\lambda$ & 0.5 & 0.7 & 0.6 & 0.75 \\
                       &  $\kappa$  & 0.033 & 0.0467 & 0.05 & 0.057 \\
                       &  f         & 15 & 15 & 12 & 13 \\ \cline{2-6}

\end{tabular}
\end{table}

We next fitted the experimental data against the model, using equations ~\ref{eq:moments} e ~\ref{eq:fractions}. These equations depend on four parameters governing the growth and division rates, two parameters fixing the first two moments of the initial size distribution, the number of intermediate tasks, Q,  and the variance of the inherited size fraction, each playing a critical role in shaping the model's behavior.  
However, it is possible to reduce the number of free parameters within the simulation: the exponent of the growth rate $\alpha$ is set to 1 to reproduce an exponential growth dynamic \cite{miotto2023determining} while both the mean and the variance of the initial size distribution can be obtained from the FSC distributions.
To take into account the effect of the stress, we assumed that parameters can vary in time. To ensure that found equations remain well defined, we opted for the introduction of a time of delay, $T_d$, such that parameters are constants before and after this time, even if their values can be different in the two-time intervals.

\begin{figure*}
    \centering
    \includegraphics[width=\textwidth]{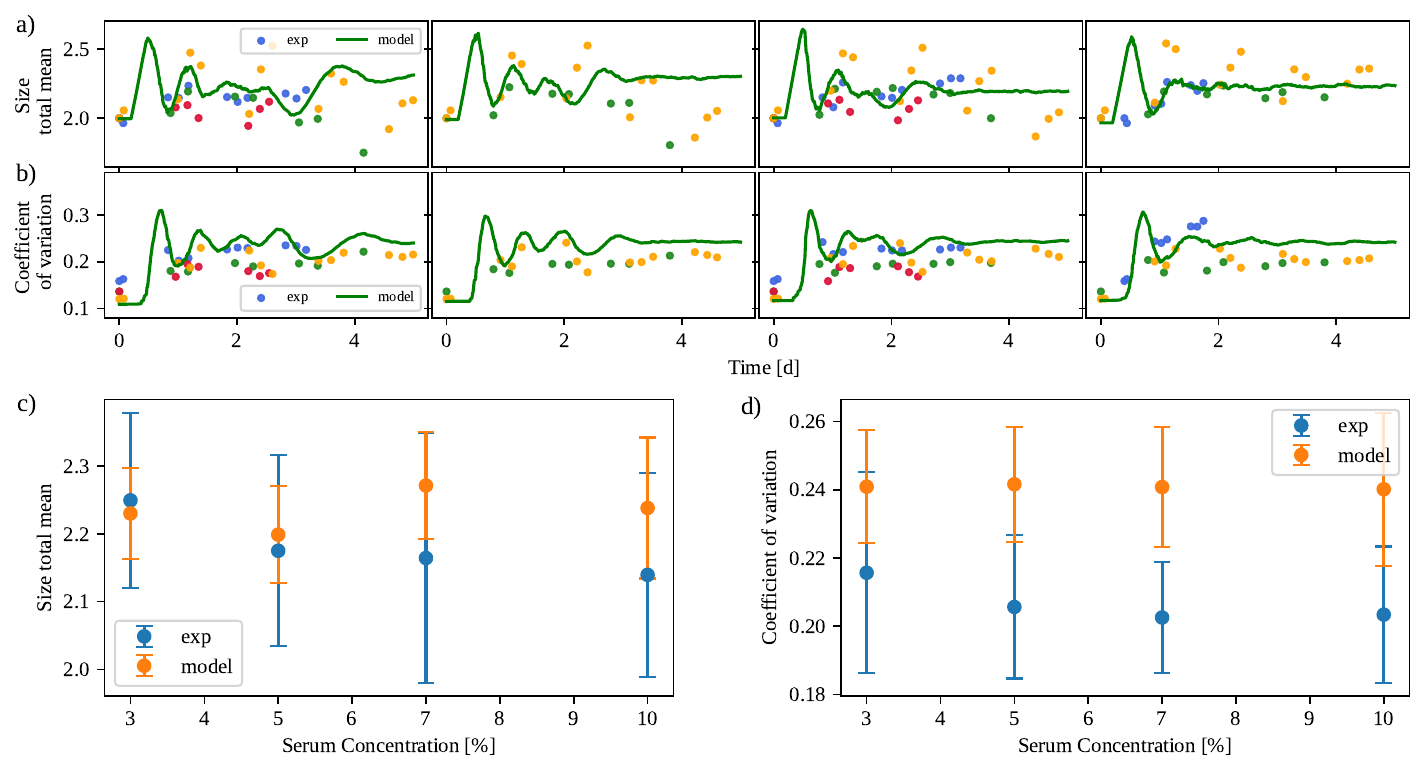}
    \caption{\textbf{Conservation of cell size distribution moments.}
    \textbf{a)}  Mean re-scaled total size given by the forward scattering measurements as a function of time (dots) and curves given by the fit of the minimal model. From left to right the serum concentration decreases (Ctrl, 7\%, 5\%, and 3\%).
    \textbf{b)}  Same as in panel a) but for the CV.   
    \textbf{c)} Mean total size and  \textbf{d)} coefficient of variation for cells in the stressed phase as a function of the serum concentration supplemented to the growing media.
    }
    \label{fig:3}
\end{figure*}

Figure~\ref{fig:2}a shows the results of the best fit of the experimental data for the cell fractions belonging to each generation as a function of time. From top to bottom, the panels refer to cells grown with gradually decreasing levels of serum.  
The best-fit parameters are reported in Table~\ref{tab:parameters}, divided into two sets: the one that best fits data before $T_d$, and the second for data after $T_d$. The values of $T_d$ as a function of the serum concentrations are shown in Figure~\ref{fig:2}b.
 It can be seen that this change is introduced earlier the lower the serum concentration in the growth medium.
We note that pre-sorting cells have been grown for at least one week in standard growing conditions (i.e. in medium with 10\% serum). Our setup equals a sudden switch to a different (more stressful) environment. This causes the cells to initially proliferate with resources taken from the previous environment.
The presence of a delay time of $\sim$2.5  for cells grown in a standard medium can be instead linked to the fact that proliferation consumes the serum as well. As a consequence, at the end of the experiment, we will have soil with serum levels that will no longer correspond to a nominal 10\% value but will be lower. The four panels in figure~\ref{fig:3}c show the proliferation curves by the level of starvation. The slope of the curve corresponds to the proliferation rate of the cell population (not to be confused with the proliferation rate of the individual cell): the steeper the curve, the higher it will be.
It can be seen that the slope is initially higher and then decreases once the lag time has passed.
This corresponds, as mentioned earlier, to a slowed-down proliferation due to the need to balance proliferation with the concentration of nutrients in the medium. Thus, the cells, even if grown in soil with low serum concentrations, will continue to proliferate.
Finally, looking at Table \ref{tab:parameters}, one can see that the parameters that change are the constants related to the growth and division rates, $\lambda$ and $\kappa$ respectively. This suggests that the cells do not change their homeostatic size control mechanism ($\beta$ remains the same), but slow down proliferation to adapt to the nutrient levels in the medium.
The $\beta$ parameter remains compatible with 5, i.e. with a sizer-like division strategy upon varying the growth conditions.

\subsection{Size distribution moments in the long time limit}

Next, we analyzed the cell size distribution moments.
Figure~\ref{fig:4}a shows the asymptotic averages of the size of the whole population for the different growth conditions.
The values of the FSC intensities, used as a proxy for cell size, were rescaled with the initial population averages. Large fluctuations in the dynamics and the attainment of a stationary level with a higher value than at the starting point can be seen. This is because the cell population is initially sorted, which leads to the initial size distribution being out of equilibrium. 
When comparing the stationary values of the mean size achieved for the four levels of starvation to which the cells were treated (Figures~\ref{fig:4}c and ~\ref{fig:4}d), an increase in the size is observed as a function of the starvation level. On the contrary, as the percentage of serum present in the growth medium decreases, the value of the asymptotic coefficient of variation remains constant, suggesting that cells tend to increase in size before dividing if the surrounding environment is nutrient-depleted. This is in line with the hypothesis that proliferation slows down in the face of increased adaptation to the growth environment. However, the population adjusts the proliferation to preserve the same level of heterogeneity, as measured by the size CV.

\subsection{Asymmetry division effects on starvation}
\begin{figure*}
    \centering
    \includegraphics[width=\textwidth]{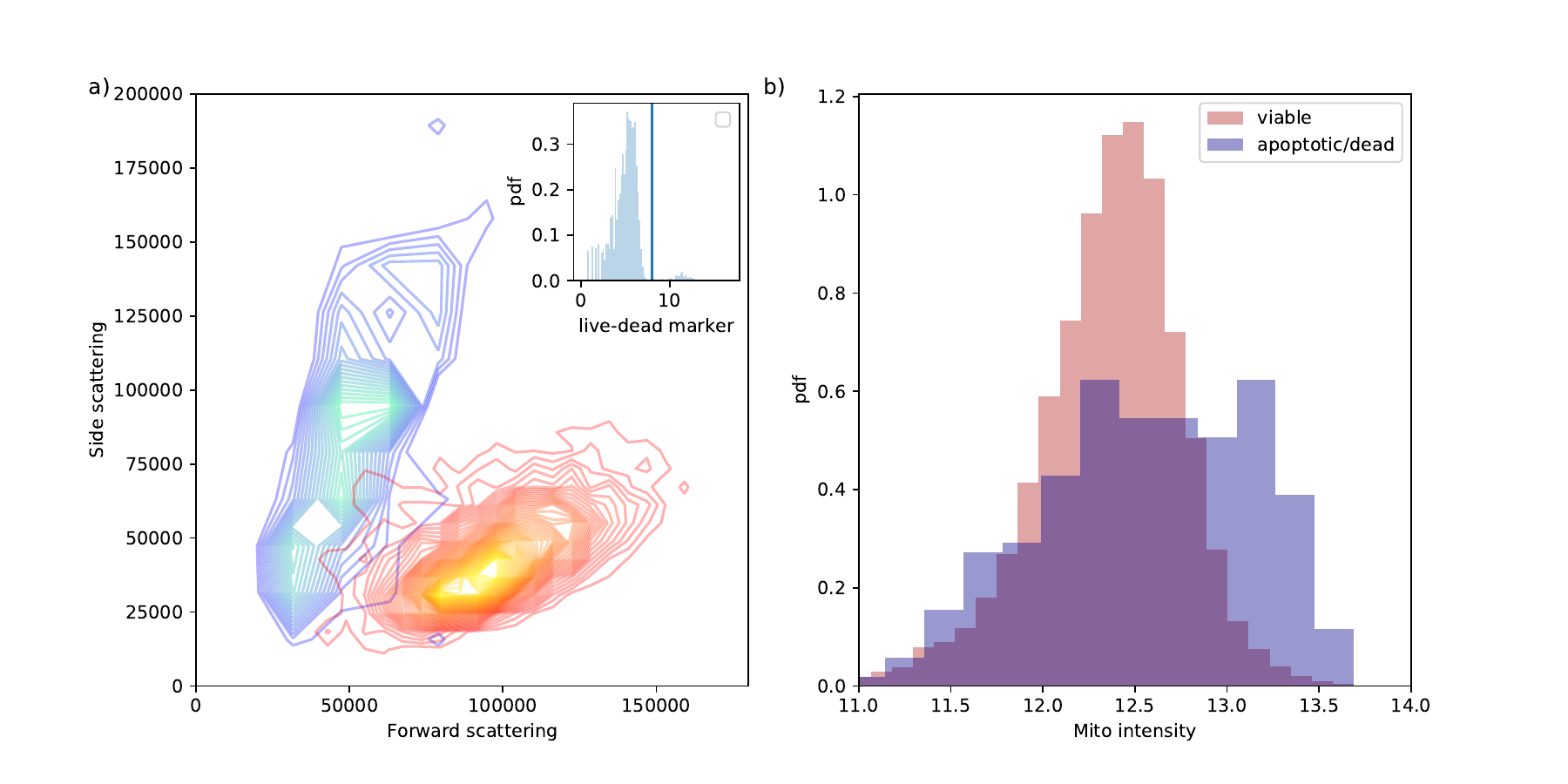}
    \caption{\textbf{Effect of phenotypic heterogeneity on stress response.} 
    \textbf{a)} Probability density function of observing viable (red) or apoptotic/dead (blue) cells as a function of their forward and side scattering intensities. Cell viability is measured by a live-dead marker FVS780, as shown in the inset and discussed in the Method section. 
    \textbf{b)} Distribution of mitochondria fluorescence intensity found in viable vs apoptotic cells belonging to the second generation for a cell population proliferating in a 3\% serum media.
    }
    \label{fig:4}
\end{figure*}

Finally, we investigated whether proliferating and non-proliferating cells exhibit differences in the manifested phenotypes. To begin with, we looked for differences in the Forward vs Side scattering plane, that are commonly used as proxies for cell size and granularity, respectively. Using  Fixable Viability Stain 780 (FVS780, BD Horizon, BD Biosciences, USA), a dye, able to discriminate viable from non-viable cells, we obtained the distribution of viable, proliferating cells  (respectively the one of non-proliferating-apoptotic cells) as those having values of marker fluorescence higher (resp. lower) than $2^7$, as shown in Figure~\ref{fig:4}a.  As for such kind of cell, the two distributions are well separated, we used this data to classify viable vs non-viable cells in all collected time points. As previous works linked serum deprivation to oxidative stress~\cite{White2020}, we looked at the mitochondria content of viable vs apoptotic cells. In particular, we used Mitotracker dye to stain cells and follow the dynamics of the fluorescence intensity as cells grow and divide. Notably, cells in the second to third generation apart from the originally stained ones, display different distributions of fluorescence. As shown in Figure~\ref{fig:4}b, viable cells have a distribution shifted toward lower values of Mitotracker than apoptotic cells.

\section{Discussions}

The ability of cancer cells to exhibit heterogeneity in their phenotypic distributions is a key factor in enhancing tumor invasiveness and drug resistance. Understanding the mechanisms behind such heterogeneity~\cite{evolat} in mammalian cells, particularly cancer cells, poses significant challenges~\cite{DagogoJack2017, Miotto2023coll}. Unlike unicellular organisms such as bacteria and yeast, mammalian cells have prolonged proliferation timescales, making it difficult to observe and analyze changes over time.
In our study, we investigated the response of leukemia cell populations to nutrient starvation using a novel protocol that combines live cell fluorescence labeling and flow cytometry with a quantitative analytical model. This approach allowed us to monitor the population proliferation and the dynamics of its phenotype distributions over time.
Leukemia cells were found to maintain a memory of their previous environment for up to one generation even under severe nutrient depletion. This memory effect suggests a form of cellular resilience and adaptability, where cells retain information about past conditions to better prepare for future challenges.
Moreover,  upon transitioning from serum-rich to serum-poor media, leukemia cells adjusted their growth and division rates to the new environment. Despite these changes, they preserved a sizer-like division strategy, where cells divide after reaching a certain target size. This indicates that the cells can modulate their behavior based on nutrient availability while maintaining a fundamental growth strategy.
 By examining the mitochondrial content of proliferating versus non-proliferating cells, we found that non-proliferating cells had a higher number of older mitochondria. This observation suggests a possible functional role for the asymmetric partitioning of aged mitochondria in leukemia cells. The presence of older mitochondria in non-proliferating cells might be linked to a strategy where cells allocate damaged or less efficient organelles to certain subsets of the population, thereby enhancing the overall fitness of the proliferating cells.

These findings in perspective may have implications for our understanding of cancer cell adaptation and heterogeneity. The ability of leukemia cells to adapt their growth and division rates based on nutrient availability, while preserving a core division strategy, highlights the complexity and plasticity of cancer cell populations. The observed memory effect and asymmetric mitochondrial partitioning add further layers to this complexity, suggesting that cancer cells use sophisticated mechanisms to survive and thrive under varying environmental conditions.
Future research should aim to dissect the molecular mechanisms underlying these adaptive behaviors. Understanding how leukemia cells retain memory of past conditions and how they regulate mitochondrial partitioning could provide new insights into cancer biology and potentially reveal novel therapeutic targets. Additionally, extending these findings to other types of cancer cells and exploring the role of different environmental stressors could help to generalize the observed phenomena and enhance our overall understanding of tumor heterogeneity and resistance.

In summary, our study sheds light on the behavior of leukemia cells in responding to nutrient starvation, emphasizing the importance of considering both growth dynamics and cellular heterogeneity in cancer research. By unraveling these mechanisms, we could move closer to understanding tumor growth and resistance.

\section{Materials and Methods}

\subsection*{Cell culture} 

E6.1 Jurkat cells were used as a cell model for proliferation study and maintained in RPMI-1640 complete culture media containing 10\% FBS, penicillin/streptomycin plus glutamine at 37 C in 5\% CO2. Upon thawing, cells were passaged once before amplification for the experiment. Cells were then harvested, counted, and washed twice in serum-free solutions and re-suspended in PBS for further staining~\cite{peruzzi2021asymmetric}. 

\subsection*{Cells fluorescent dye labeling}

To track cell proliferation by dye dilution establishing the daughter progeny of a completed cell cycle, cells were stained with CellTrace\textsuperscript{\texttrademark} Violet stain (CTV, C34557, Life Technologies, Paisley, UK), typically used to monitor multiple cell generations, and MitoTracker\textsuperscript{\textregistered} Deep Red 633 (M22426, Molecular Probes, Eugene, USA). To determine cell viability, prior to dye staining, the collected cells were counted with the hemocytometer using the dye exclusion test of Trypan Blue solution, an impermeable dye not taken up by viable cells. To reduce the time that cells are incubated with different dyes, we optimized the protocol performing the simultaneous staining of CTV and Mitotracker Deep Red.
For the dyes co-staining, highly viable  $20\times10^6$ cells were incubated in a 2ml solution of PBS containing both CTV (1/1000 dilution according to the manufacturer’s instruction) and Mitotracker Deep Red (used at a final concentration of 200nM) for 25 min at room temperature (RT) mixing every 10min to ensure homogeneous cell labeling. Afterward, complete media was added to the cell suspension for an additional 5 min incubation before the final washing in PBS.
In the experiments performed to evaluate viability, cells were stained with FVS780 used 1/1000 in PBS w/o salt for 10 minutes at room temperature in the dark. Upon extensive washing, cells were run at the flow cytometer and plotted with the FSC parameter to establish live cell distribution.

\subsection*{Cell sorting}

Jurkat cells labeled with dyes were sorted using a FACSAriaIII (Becton Dickinson, BD Biosciences, USA) equipped with Near UV 375nm, 488nm, 561nm, and 633nm lasers and FACSDiva software (BD Biosciences version 6.1.3). Data were analyzed using FlowJo software (Tree Star, version 9.3.2 and 10.10.0). Briefly, cells were first gated on single cells, by doublets exclusion with morphology parameters, both side and forward scatter, area versus width (A versus W). The unstained sample was used to set the background fluorescence for each channel. For each fluorochrome, a sorting gate was set around the max peak of fluorescence of the dye distribution~\cite{Filby2015}. In this way, the collected cells were enriched for the highest fluorescence intensity for the markers used. Following isolation, an aliquot of the sorted cells was analyzed with the same instrument to determine the post-sorting purity and population width, resulting in an enrichment $>99\%$ for each sample.

\subsection*{Serum-starvation and Time course kinetic for dye dilution assessment}
The sorted cell population was seeded in four wells of a 6-well plate (BD Falcon), $\sim0.3\times10^6$ cells/well, and kept in culture for up to 72/96 hours. The four wells differ in the concentration of serum in the growth medium (RPMI-1640 complete culture media containing penicillin/streptomycin plus glutamine). Specifically, four media were prepared, each with a different serum concentration: Ctrl (10\% addition of FBS), 7\%, 5\%, and 3\%. 

To monitor multiple cell division, an aliquot of the cells in culture from each well was analyzed three times a day, every day, for the duration of the experiment to measure the fluorescence intensity of CTV dye by the LSRFortessa flow cytometer.
To set the time zero of the kinetic, prior culturing, a tiny aliquot of the collected cells was analyzed immediately after sorting at the flow cytometer. The unstained sample was used to set the background fluorescence as described above. Every time that an aliquot of cells was collected for analysis, the same volume of fresh media with the specific serum concentration was replaced in the respective culture well.

\subsection*{Expectation-Maximization and the Gaussian Mixture Model}

We used the Expectation-Maximization (EM) algorithm
to detect the clusters
in Gaussian Mixture Models~\cite{Dempster1977}. The EM algorithm is composed of two steps the
Expectation (E) step and the Maximization (M) step.
In the E-step, for each data point $\bf{f}$,
we used our current guess of $\pi_g $, $\mu_g$, and $\sigma_g$,
to estimate the posterior probability that each cell belongs to generation $g$
given that its fluorescence intensity measure
as $\bf{f}$, $\gamma_g=\mathrm{P}(g|\bf{f})$.
In the M-step, we use the fact that the gradient of the log-likelihood of 
$p(\bf{f}_i)$ for $\pi_g $, $\mu_g$, and $\sigma_g$
can be computed. Consequently, the expression of
the optimal value of $\pi_g $, $\mu_g$, and $\sigma_g$ is dependent on $\gamma_g$.
It is shown that under, certain smoothness conditions the iterative computation
of E-stem and M-step leads us to the locally optimal estimate of the parameters
$\pi_g $, $\mu_g$, and $\sigma_g$, and returns
the posterior probability $\gamma_g$ which weights how much each point belongs to one of the clusters.
Here, we used this model
to perform cluster analysis and detect the peaks which correspond to different
generations. Then, we estimated $\pi_g$, $\mathrm{E}[f_g]$, and $\mathrm{Var}[f_g]$ from these clusters.
Note that assuming a Gaussian Mixture model assures the less biased fitting procedure to get the second moment of the underlining distributions according to the maximum entropy principle~\cite{ecolat, tolomeo}.

\subsection*{Gillespie simulation}

To validate the mathematical model that was formulated, stochastic simulations of the
growth and dividing cell population were carried out. Note that, through simulation, we can also know the birth and division sizes and can therefore compare the trends of  $\Delta$  Vs $<s_b>$.

In particular, simulations were performed starting from $N=1000$ initial cells, having initial size randomly sampled from a normal distribution of mean $\mu_s$ and variance $\sigma^2_s$.

For each cell, a division time is extracted from the probability distribution $P(t_d)$ via inverse transform sampling.  For the considered system, $P(t_d)$ is given by~\cite{NietoAcuna2019}:
\beq
P(t_d) = 1 - \exp \Big( - \int_0^{t_d} dt h(s)\Big)
\eeq

Upon division, each cell is split into two new daughter cells, each inheriting a fraction $p$ and $(1-p)$ of the mother size, respectively.


\section*{Data Availability}

The data that support the findings of this study are available from the corresponding
author upon reasonable request.

\section*{Code Availability}

All codes used to produce the findings of this study are available from the corresponding author upon request.
The code for the Gaussian Mixture algorithm is available at \href{https://github.com/ggosti/fcMGM}{https://github.com/ggosti/fcGMM}.\\

\section*{Author contributions statement}

M.M. conceived research;  G.R. contributed additional ideas; S.S. and G.P. performed experiments;  S.S. and G.G. analyzed data and performed numerical simulations and statistical analysis; M.M. performed analytical calculations; all authors analyzed results; S.S. and M.M. wrote the paper; all authors revised the paper.

\section*{Competing Interests}
The authors declare no competing interests.

\section*{Acknowledgements}
This research was partially funded by grants from ERC-2019-Synergy Grant (ASTRA, n. 855923); EIC-2022-PathfinderOpen (ivBM-4PAP, n. 101098989); Project `National Center for Gene Therapy and Drugs based on RNA Technology' (CN00000041) financed by NextGeneration EU PNRR MUR—M4C2—Action 1.4—Call `Potenziamento strutture di ricerca e creazione di campioni nazionali di R\&S' (CUP J33C22001130001).

\bibliographystyle{unsrt}
\bibliography{mybib}

\begin{thebibliography}{10}

\bibitem{hoffmann1997extreme}
A.A. Hoffmann and P.A. Parsons.
\newblock {\em Extreme Environmental Change and Evolution}.
\newblock Extreme Environmental Change and Evolution. Cambridge University
  Press, 1997.

\bibitem{expexp}
Andrea De~Martino, Thomas Gueudr{\'e}, and Mattia Miotto.
\newblock Exploration-exploitation tradeoffs dictate the optimal distributions
  of phenotypes for populations subject to fitness fluctuations.
\newblock {\em Physical Review E}, 99(1):012417, 2019.

\bibitem{Miotto2020diff}
Mattia Miotto, Lorenzo Di~Rienzo, Pietro Corsi, Giancarlo Ruocco, Domenico
  Raimondo, and Edoardo Milanetti.
\newblock Simulated epidemics in 3d protein structures to detect functional
  properties.
\newblock {\em Journal of Chemical Information and Modeling},
  60(3):1884–1891, February 2020.

\bibitem{Wilson2010}
Wayne~A. Wilson, Peter~J. Roach, Manuel Montero, Edurne Baroja-Fernández,
  Francisco~José Muñoz, Gustavo Eydallin, Alejandro~M. Viale, and Javier
  Pozueta-Romero.
\newblock Regulation of glycogen metabolism in yeast and bacteria.
\newblock {\em FEMS Microbiology Reviews}, 34(6):952–985, November 2010.

\bibitem{Sekar2020}
Karthik Sekar, Stephanie~M. Linker, Jen Nguyen, Alix Gr\"{u}nhagen, Roman
  Stocker, and Uwe Sauer.
\newblock Bacterial glycogen provides short-term benefits in changing
  environments.
\newblock {\em Applied and Environmental Microbiology}, 86(9), April 2020.

\bibitem{EnricoBena2021}
Chiara Enrico~Bena, Marco Del~Giudice, Alice Grob, Thomas Gueudré, Mattia
  Miotto, Dimitra Gialama, Matteo Osella, Emilia Turco, Francesca Ceroni,
  Andrea De~Martino, and Carla Bosia.
\newblock Initial cell density encodes proliferative potential in cancer cell
  populations.
\newblock {\em Scientific Reports}, 11(1), March 2021.

\bibitem{Kellogg2022}
Douglas~R. Kellogg and Petra~Anne Levin.
\newblock Nutrient availability as an arbiter of cell size.
\newblock {\em Trends in Cell Biology}, 32(11):908–919, November 2022.

\bibitem{Kafri2013}
Ran Kafri, Jason Levy, Miriam~B. Ginzberg, Seungeun Oh, Galit Lahav, and
  Marc~W. Kirschner.
\newblock Dynamics extracted from fixed cells reveal feedback linking cell
  growth to cell cycle.
\newblock {\em Nature}, 494(7438):480–483, February 2013.

\bibitem{Sung2013}
Yongjin Sung, Amit Tzur, Seungeun Oh, Wonshik Choi, Victor Li, Ramachandra~R.
  Dasari, Zahid Yaqoob, and Marc~W. Kirschner.
\newblock Size homeostasis in adherent cells studied by synthetic phase
  microscopy.
\newblock {\em Proceedings of the National Academy of Sciences},
  110(41):16687–16692, September 2013.

\bibitem{peruzzi2021asymmetric}
Giovanna Peruzzi, Mattia Miotto, Roberta Maggio, Giancarlo Ruocco, and Giorgio
  Gosti.
\newblock Asymmetric binomial statistics explains organelle partitioning
  variance in cancer cell proliferation.
\newblock {\em Communications Physics}, 4(1):188, 2021.

\bibitem{miotto2023determining}
Mattia Miotto, Simone Scalise, Marco Leonetti, Giancarlo Ruocco, Giovanna
  Peruzzi, and Giorgio Gosti.
\newblock A size-dependent division strategy accounts for leukemia cell size
  heterogeneity.
\newblock {\em Communications Physics}, 7(1), July 2024.

\bibitem{Osella2014}
Matteo Osella, Eileen Nugent, and Marco~Cosentino Lagomarsino.
\newblock Concerted control of escherichia coli cell division.
\newblock {\em Proceedings of the National Academy of Sciences},
  111(9):3431--3435, February 2014.

\bibitem{NietoAcuna2019}
Cesar~Augusto Nieto-Acuna, Cesar~Augusto Vargas-Garcia, Abhyudai Singh, and
  Juan~Manuel Pedraza.
\newblock Efficient computation of stochastic cell-size transient dynamics.
\newblock {\em {BMC} Bioinformatics}, 20(S23), December 2019.

\bibitem{Totis2021}
Niccolo Totis, Cesar Nieto, Armin Kuper, Cesar Vargas-Garcia, Abhyudai Singh,
  and Steffen Waldherr.
\newblock A population-based approach to study the effects of growth and
  division rates on the dynamics of cell size statistics.
\newblock {\em {IEEE} Control Systems Letters}, 5(2):725--730, April 2021.

\bibitem{Jia2021}
Chen Jia, Abhyudai Singh, and Ramon Grima.
\newblock Cell size distribution of lineage data: Analytic results and
  parameter inference.
\newblock {\em {iScience}}, 24(3):102220, March 2021.

\bibitem{Nieto2020}
C\'esar Nieto, Juan Arias-Castro, Carlos S\'anchez, C\'esar Vargas-Garc\'{\i}a,
  and Juan~Manuel Pedraza.
\newblock Unification of cell division control strategies through continuous
  rate models.
\newblock {\em Phys. Rev. E}, 101:022401, Feb 2020.

\bibitem{Yates2017}
Christian~A. Yates, Matthew~J. Ford, and Richard~L. Mort.
\newblock A multi-stage representation of cell proliferation as a markov
  process.
\newblock {\em Bulletin of Mathematical Biology}, 79(12):2905--2928, October
  2017.

\bibitem{White2020}
ElShaddai~Z. White, Nakea~M. Pennant, Jada~R. Carter, Ohuod Hawsawi, Valerie
  Odero-Marah, and Cimona~V. Hinton.
\newblock Serum deprivation initiates adaptation and survival to oxidative
  stress in prostate cancer cells.
\newblock {\em Scientific Reports}, 10(1), July 2020.

\bibitem{evolat}
Mattia Miotto and Lorenzo Monacelli.
\newblock Genome heterogeneity drives the evolution of species.
\newblock {\em Physical Review Research}, 2(4), October 2020.

\bibitem{DagogoJack2017}
Ibiayi Dagogo-Jack and Alice~T. Shaw.
\newblock Tumour heterogeneity and resistance to cancer therapies.
\newblock {\em Nature Reviews Clinical Oncology}, 15(2):81–94, November 2017.

\bibitem{Miotto2023coll}
Mattia Miotto, Maria Rosito, Matteo Paoluzzi, Valeria de~Turris, Viola Folli,
  Marco Leonetti, Giancarlo Ruocco, Alessandro Rosa, and Giorgio Gosti.
\newblock Collective behavior and self-organization in neural rosette
  morphogenesis.
\newblock {\em Frontiers in Cell and Developmental Biology}, 11, August 2023.

\bibitem{Filby2015}
Andrew Filby, Julfa Begum, Marwa Jalal, and William Day.
\newblock {Appraising the suitability of succinimidyl and lipophilic
  fluorescent dyes to track proliferation in non-quiescent cells by dye
  dilution}.
\newblock {\em Methods}, 82:29--37, jul 2015.

\bibitem{Dempster1977}
A.~P. Dempster, N.~M. Laird, and D.~B. Rubin.
\newblock Maximum likelihood from incomplete data via the em algorithm.
\newblock {\em Journal of the Royal Statistical Society. Series B
  (Methodological)}, 39(1):1--38, 1977.

\bibitem{ecolat}
Mattia Miotto and Lorenzo Monacelli.
\newblock Entropy evaluation sheds light on ecosystem complexity.
\newblock {\em Physical Review E}, 98(4), October 2018.

\bibitem{tolomeo}
Mattia Miotto and Lorenzo Monacelli.
\newblock Tolomeo, a novel machine learning algorithm to measure information
  and order in correlated networks and predict their state.
\newblock {\em Entropy}, 23(9):1138, August 2021.

\end{thebibliography}

\end{document}